# Complementarity principle on human longevity

**Abstract**

In recent we introduced, developed and established a new concept, model, methodology and principle for studying human longevity in terms of demographic basis. We call the new model the "Weon model", which is a general model modified from the Weibull model with an age-dependent shape parameter to describe human survival and mortality curves. We demonstrate the application of the Weon model to the mortality dynamics and the mathematical limit of longevity (the mortality rate to be mathematically zero, implying a maximum longevity) in the Section I. The mathematical limit of longevity can be induced by the mortality dynamics in nature. As a result, we put forward the complementarity principle, which explains the recent paradoxical trends that the mathematical limit decreases as the longevity increases, in the Section II. Our findings suggest that the human longevity can be limited by the complementarity principle.

**Acknowledgements**

The author is grateful to Dr. James W. Vaupel and Dr. S. Jay Olshansky for stimulating comments, and to Dr. J. H. Je for useful discussions. The author also thanks to the Human Mortality Database (Dr. John R. Wilmoth, as a director, in The University of California, Berkeley and Dr. Vladimir Shkolnikov, as a co-director, in The Max Planck Institute for Demographic Research) for allowing anyone to access the demographic data for research.

# Section I. Mortality dynamics and mathematical limit

**Summary**

In this Section, we wish to show that the mortality dynamics is a consequence of the bending of the shape parameter at old ages. This investigation is based upon the Weon model (the Weibull model with an age-dependent shape parameter) for human survival and mortality curves. According to the Weon model, we are well able to describe the mortality decrease after the mortality plateau, including the mortality deceleration. Furthermore, we are able to simply define the mathematical limit of longevity by the mortality decrease. From the demographic analysis of the historical trends in Switzerland (1876-2001) and Sweden (1861-2001), and the most recent trends in the other eleven developed countries (1996-2001), we confirm that the bending of the shape parameter after characteristic life is correlated with the mortality deceleration (or decrease). As a consequence, this bending of the shape parameters and the mortality deceleration is associated with the mathematical limit on longevity. These results suggest that the mathematical limit of longevity can be induced by the mortality deceleration (or decrease) in nature.

# 1. Introduction

Fundamental studies of the aging process have lately attracted the interest of researchers in a variety of disciplines, linking ideas and theories from such diverse fields as biochemistry to mathematics (Weitz and Fraser 2001). The way to characterize aging is to plot the increase in mortality rate with chronological age. The mortality rate is the probability that an individual who is alive at a particular age will die during the following age interval. The mortality rate can also be represented as the fraction of the population surviving to a particular age (or the survival rate).

The fundamental law of population dynamics is the Gompertz law (Gompertz 1825), in which the human mortality rate increases roughly exponentially with increasing age at senescence. The Gompertz model is most commonly employed to compare mortality rates between different populations (Penna and Stauffer 1996). However, no mathematical model so far, including the Gompertz model, has been suggested that can perfectly approximate the development of the mortality rate over the total life span (Kowald 1999). Particularly in modern research findings, it seems to be obvious that the mortality rate does not increase according to the Gompertz model at the highest ages (Vaupel 1997; Robine and Vaupel 2002), and this deviation from the Gompertz model is a great puzzle to demographers, biologists and gerontologists. There are two standard hypotheses that have been put forward to explain this phenomenon: the individual-risk hypothesis and the heterogeneity hypothesis (Higgins 2003). This puzzle needs to be resolved.

There is strong evidence from many developed countries that the rate of increase in mortality decelerates at high ages. However, many of the traditional mathematical models (for instance, the Gompertz, Weibull, Heligman & Pollard, Kannisto, Quadratic and Logistic models) for the mortality rate provide poor fits to empirical population data at the highest ages (Thatcher, Kannisto and Vaupel 1998; Yi and Vaupel 2003). We have recently found a useful model derived from the Weibull model with an age-dependent shape parameter to describe the human survival and mortality curves (Weon 2004a, 2004b).

The model suggests that the mortality deceleration (or decrease) at old ages is a consequence of the bending of the shape parameter (Weon 2004a). We are able to apply the model to describe the mortality deceleration (or decrease) at older ages. In this investigation, we wish to demonstrate the mortality deceleration (or decrease) by the bending of the shape parameter through the demographic analysis of the historical trends in Switzerland (1876-2001) and Sweden (1861-2001), and the most recent trends in the other eleven developed countries (1996-2001). In fact, the mortality rate decelerates at higher ages, reaching perhaps a maximum or ceiling around age 110 (Vaupel et al. 1998; Helfand and Inouye 2002). We will show that the new model enables us to describe the mortality dynamics at the highest ages.

**2. Weon model**

We have put forward a general expression for human survival and mortality rate in our previous papers (Weon 2004a, 2004b). It has been recently discovered that human survival and mortality curves are well described by the following new mathematical model, derived from the Weibull survival function and it is simply described by two parameters, the age-dependent shape parameter and characteristic life, as follows,

$$S = \exp(-(t/\alpha)^{\beta(t)}) \qquad (1)$$

where $S$ denotes the survival probability of surviving to age $t$, $\alpha$ denotes characteristic life and $\beta(t)$ denotes the shape parameter as a function of age. The original idea was obtained as follows: typical human survival curves show i) a rapid decrease in survival in the first few years of life and ii) a relatively steady decrease and then an abrupt decrease near death thereafter (see Fig. 1 in Weon 2004a; Azbel 2002). Interestingly, the former behaviour resembles the Weibull survival function with $\beta < 1$ and the latter behaviour seems to follow the case of $\beta \gg 1$. With this in mind, it could be assumed that shape parameter is a function of age. The new model is completely different from the Weibull model in terms of the 'age dependence of the shape parameter'. It is especially noted that the shape parameter '$\beta(t) = \ln(-\ln S)/\ln(t/\alpha)$' can indicate a 'rectangularity' of the survival curve. The reason for this is that as the value of the shape parameter becomes a high value, the shape of the survival curve approaches a further rectangular shape. We will call the new model the 'Weon model' hereafter.

The Weibull model for technical devices has a constant shape parameter, whereas the Weon model for humans has an age-dependent shape parameter; $S = \exp(-(t/\alpha)^{\beta})$ for technical devices (the Weibull model) and $S = \exp(-(t/\alpha)^{\beta(t)})$ for humans (the Weon model). In the Weon model, the shape parameter for humans, '$\beta(t) = \ln(-\ln S)/\ln(t/\alpha)$', is a function of age $t$. Therefore, the mortality function is described by the mathematical relationship with the survival function ($\mu = -d \ln S/dt$) and is in general as follows,

$$\mu = \frac{d}{dt}((t/\alpha)^{\beta(t)}) \text{ or } \mu = (t/\alpha)^{\beta(t)} \times [\frac{\beta(t)}{t} + \ln(t/\alpha) \times \frac{d\beta(t)}{dt}] \tag{2}$$

where $\mu$ is the mortality rate (denoted as the hazard rate or the force of mortality), meaning the relative rate for the survival function decrease. Obviously, mortality trends should be directly associated with shape parameter trends. It is noteworthy that the 'age dependence of the shape parameter' intrinsically makes the mortality function complex and difficult for modeling.

Conveniently, the value of characteristic life ($\alpha$) is always found at the duration for survival to be '$\exp(-1)$'; this is known as the characteristic life. This feature gives the advantage of looking for the value of $\alpha$ simply by a graphical analysis of the survival curve. In turn, with the observed value of $\alpha$, we can plot 'rectangularity' with age by the mathematical equivalence of '$\beta(t) = \ln(-\ln S)/\ln(t/\alpha)$'. If $\beta(t)$ is not constant with age, this obviously implies that '$\beta(t)$ is a function of age'. On the other hand, $\beta(t)$ mathematically approaches infinity as the age $t$ approaches the value of $\alpha$ or the denominator '$\ln(t/\alpha)$' approaches zero. This feature of $\beta(t)$ can leave 'traces of $\alpha$' in the plot of $\beta(t)$, so we can observe variations of $\beta(t)$ and $\alpha$ at once in the plot of the shape parameters. If $\beta(t)$ (except for the mathematical singularity (traces of $\alpha$)) can be expressed by an adequate mathematical function, the survival and mortality functions can be calculated by the mathematically expressed $\beta(t)$. Only two parameters, $\beta(t)$ and $\alpha$, determine the survival and mortality functions. In empirical practice, we could use a linear expression for the mature phase (middle age) and a quadratic expression for senescence phase (from characteristic life to maximum age):

$\beta(t) = \beta_0 + \beta_1 t + \beta_2 t^2 + ...$, where the associated coefficients were determined by a regression analysis in the plot of shape parameter curve. And thus, the derivative of $\beta(t)$ was obtained as follows: $d\beta(t)/dt = \beta_1 + 2\beta_2 t + ...$, which is the important finding in the previous papers (Weon 2004a, 2004b), that is, the shape parameter for humans is a function of age.

## 3. An approximate relationship between the mortality rate and the shape parameter

We are able to observe a certain degree of universality for the linear progression of the shape parameter for ages earlier than characteristic life in many modern developed countries. Figure 1A shows that in thirteen developed countries (Austria 1999, Canada 1996, Denmark 2000, England & Wales 1998, Finland 2000, Italy 1999, Germany 1999, Japan 1999, Netherlands 1999, Norway 2000, Sweden 2001, Switzerland 2001 and USA 1999), the shape parameter for ages 0 up to 80 was estimated to be $\beta(t) = 1.44578 + 0.08581t$. It was shown in the previous paper (Weon 2004a), that these countries have a characteristic life of more than 80 years. Using this universality of $\beta(t)$, we could simulate the relationship of the mortality deceleration and the shape parameter in the following way.

For our analysis, we separately evaluate the impact of the mathematical terms on the mortality rate as follows,

$$\mu = A \times B; \text{ where } A = (t/\alpha)^{\beta(t)} \text{ and } B = [\frac{\beta(t)}{t} + \ln(t/\alpha) \times \frac{d\beta(t)}{dt}] \qquad (3)$$

In Fig. 1B, since the term $B$ is almost invariant after adulthood (~20+), the result is that we could obtain the approximate relationship of the mortality rate (log-scale) and the shape parameter (linear-scale) after adulthood.

$$\ln \mu \propto \beta(t) \tag{4}$$

This relationship implies that the mortality rate would follow the Gompertz law (exponential growth in the mortality rate with age), when the shape parameter is linearly proportional to age, but the mortality rate would deviate from the Gompertz law when the shape parameter is bent (non-linear) with age.

Furthermore, the age dependence of the shape parameter, $\beta(t)$, for population dynamics may be a general law that includes the Gompertz model and the Weibull model: i) the Gompertz model when $\beta(t)$ is linearly proportional to age and ii) the Weibull model when $\beta(t)$ is a constant. The Gompertz model (Gompertz 1825) and the Weibull model (Weibull 1951) are the most generally used models at present (Gavrilov and Gavrilova 2001). Interestingly, the Gompertz model is more commonly used to describe biological systems, whereas the Weibull model is more commonly applicable to technical devices (Gavrilov and Gavrilova 2001). Particularly for aging patterns, it is the shape parameter that distinguishes humans from technical devices. It seems to show the difference between humans and technical devices in terms of '*robustness*'. The fundamental difference for robustness between biological systems and technical devices is obvious (Gavrilov and Gavrilova 2001). In the previous papers (Weon 2004a, 2004b), the age-dependent shape parameter is changed from approximately 0.5 to 10 with age for the typical human survival curves. This feature is in great contrast to technical devices typically having a constant shape parameter (Nelson 1990). We attribute the age dependence of the shape parameter to the resistance to aging, which can be described as a struggle to extend the survival probability by increasing the shape parameter, or the *homeostasis* and the *adaptation* of biological systems, which must have the homeostasis and the adaptation to maintain stability and to survive (Weon 2004a).

For humans over much of the age range, the Gompertz model still gives an excellent approximation. At the high ages, however, the law does not apply very well (Thatcher, Kannisto and Vaupel 1998). In the Weon model, a linear expression for the shape parameter is appropriate before characteristic life and a quadratic expression is appropriate after characteristic life for the modern demographic curves (Weon 2004a). The age-dependent shape parameter in the Weon model can be seen as a measure of the deviation from the Gompertz law at senescence and the mortality deceleration at high ages.

## 4. A mathematical limit of longevity by mortality decrease

In order to model the mortality rate after characteristic life (or at the high ages), we should model the shape parameter accurately. For example in Fig. 2A, we plotted $\beta(t)$ after the characteristic life in Switzerland (1876-2001); sometimes we had to omit several points close to the characteristic life to ensure the quality of the regression analysis. Also, we accomplished the regression analysis to confirm the best mathematical model for $\beta(t)$. In this case, we used a quadratic expression for $\beta(t)$. In this way, in Fig. 2B, we examined the demographic data from Sweden (1861-2001). The detailed results of the regression analysis are recorded in Table 1 (Switzerland) and Table 2 (Sweden).

The mathematical model for $\beta(t)$ enables us to model the mortality rate after characteristic life. We are able to see the example of modeling $\mu$ through modeling $\beta(t)$ for Switzerland (2002) and Sweden (2002) in Fig. 3A and 3B. It is especially noted that in Fig. 3B, the mortality rate shows the decrease after a plateau around ages 110-115 and this shows the emergence of the mathematical limit around ages 120-130.

The following way is convenient to evaluate the mathematical limit. The mortality rate should be mathematically positive ($\mu > 0$); since the term $A$ is always positive ($A > 0$), thus the term $B$ should be positive ($B > 0$) in the Eq. (3). Therefore, the criterion for the mathematical limit of longevity, implying the ultimate limit of longevity which is able to be determined by the mortality dynamics, can be given by,

$$\frac{d\beta(t)}{dt} > -\frac{\beta(t)}{t \ln(t/\alpha)} \qquad (5)$$

For analysis, the left term denotes the term $C$ and the right term denotes the term $D$. In order to evaluate the mathematical limit of longevity, the age-dependent shape parameter after characteristic life must be accurately modeled. In the previous paper (Weon 2004a), we successfully used the quadratic expression ($\beta(t) = \beta_0 + \beta_1 t + \beta_2 t^2$) for the estimation of the shape parameter after characteristic life in the case of the years of 1876-2001 in Switzerland. Interestingly, the quadratic coefficient ($\beta_2$) supports the bending of the shape parameter, which would be directly associated with the mortality deceleration as discussed above. Furthermore, the quadratic coefficient ($\beta_2$) is important to evaluate the mathematical limit of longevity, since it determines the slope with age in the derivative ($\beta_1 + 2\beta_2 t$) of the quadratic expression of the shape parameter or in the term $C$.

## 5. Demographic evidence in Switzerland and Sweden

Demographic data, the period life tables (all sexes, 1x1) for the years of 1876-2001 in Switzerland and for the years of 1861-2001 in Sweden, were taken from the Human Mortality Database (available at http://www.mortality.org). We calculated the trend lines for the shape parameters after characteristic life in Table 1 (Switzerland) and Table 2 (Sweden) by the quadratic expression for the shape parameter. We could see that the estimated quadratic expression for the shape parameter is very accurate according to the high value of the coefficient of determinant, $r^2$, and the extremely low $P$-value (<0.0001).

In order to evaluate the mathematical limits of recent decades in Switzerland and Sweden, we separately calculate the term $C$ and the term $D$, and then plot the mathematical terms as a function of age in Fig. 4A (Switzerland) and 4B (Sweden). The term $C$ decreases with increasing age, whereas the term $D$ increases with increasing age. Interestingly, the term $C$ tends to become steeper over time in recent decades. The slope of the term $C$ is important to evaluate the mathematical limit of longevity. The most significant observation is that the mathematical limit of longevity becomes shorter as the slope of the term $C$ becomes steeper in Fig. 4A and 4B. This phenomenon, as shown in Fig. 5, indicates that the bending of the shape parameter (or the quadratic coefficient) is associated with the mathematical limit of longevity, which is due to the mortality decrease at the highest ages. The slope of the term $C$ or the quadratic coefficient ($\beta_2$) tends to increase in recent decades as shown in Fig. 4A and 4B, and in Table 1 and Table 2.

**6. Demographic evidence in the other eleven countries**

The findings from Switzerland and Sweden are very interesting. We wish to expand our investigation to the other eleven developed countries (Austria 1999, Canada 1996, Denmark 2000, England & Wales 1998, Finland 2000, Italy 1999, Germany 1999, Japan 1999, Netherlands 1999, Norway 2000, Sweden 2001, Switzerland 2001 and USA 1999) as the most recent trends. Demographic data, the period life tables (all sexes, 1x1) for the most recent years between 1996-2001 in the eleven developed countries, were taken from the Human Mortality Database (available at http://www.mortality.org). We calculated the trend lines for the shape parameters after characteristic life in Table 3. We could see that the estimated quadratic expression for the shape parameter is very accurate according to the high value of the coefficient of the determinant, $r^2$, and the extremely low *P*-value (<0.0001).

We accomplished the similar examinations in Fig. 6, 7, and 8. We found out that the quadratic expression for $\beta(t)$ is appropriate after characteristic life in Fig. 6 for the eleven countries. We evaluated the mathematical limits of longevity in Fig. 7. Finally, we could obtain the similar phenomenon between the mathematical limits and the quadratic coefficient (the bending of the shape parameter) in Fig. 8. These results confirm that the bending of the shape parameter after characteristic life is correlated with the mortality deceleration (or decrease), and consequently, is associated with the mathematical limit of longevity. The findings suggest that the mathematical limit of longevity can de induced by the mortality deceleration (or decrease) in nature.

**7. Discussion**

Vaupel et al. (1998) has suggested that the human mortality rates could decrease after having reached a maximum (a plateau or a ceiling of mortality). The results reported by Robine and Vaupel (2002) strongly support the finding that the mortality rate does not increase according to the Gompertz curve at the highest ages and the results are consistent with a plateau around ages 110-115, and their earlier study suggests that mortality may fall after age 115. In humans, the deceleration in the mortality rate is not seen until after 80 years of age, but a clear deviation from the predicted Gompertz model of exponential increases in mortality is observed. If individuals such as Jeanne Calment, who lived to the ripe old age of 122 years and 164 days, are included, then mortality rates are seen to decrease with age after 110 years of age (Helfand and Inouye 2002). The Weon model is well able to describe the mortality decrease after the mortality plateau, including the mortality deceleration, and further, the model predicts the mathematical limit of longevity by the mortality decrease. It may be arguable to evaluate the mathematical limit of longevity by the mortality decrease based on this assumption, which fit the available data from characteristic life (around age 80) up to approaching age 110, and the Weon model will continue to apply up to older age. However, the Weon model for mortality rates from the estimation of the shape parameter is very reliable in this investigation, so we believe that such evaluations are reasonable.

Yet to our surprise, the trend of the mathematical limit of longevity in recent decades seems to be contrary to all our knowledge. We should note that the characteristic life in both of Switzerland and Sweden has increased constantly for more than a century in Fig. 9, which is consistent with other literature (for instance, Wilmoth et al. 2000; Oeppen and Vaupel 2002; Robine and Vaupel 2002). On the contrary, the mathematical limits of longevity by the mortality decrease seem to tend to decrease in recent decades. Furthermore, the mathematical limit of longevity in the thirteen developed countries approaches approximately 120 to 130 years in recent decades in Fig. 4 and 5, and Fig. 7 and 8. Interestingly, the bending of the shape parameter tends to be more intensive with increasing characteristic life in the previous paper (see Fig. 6 in Weon 2004a). This suggests that there probably is an inherent correlation between the mortality dynamics and the increase of characteristic life (or longevity tendency). The mathematical limit of longevity may be associated with the fundamental principle of human longevity. We intend to investigate this question in the Section II.

## 8. Conclusions

The age-dependent shape parameter can be a measure of the deviation from the Gompertz law at senescence or the mortality deceleration (or decrease) at the highest ages. According to the Weon model, we are well able to describe the mortality decrease after the mortality plateau, including the mortality deceleration. Furthermore, we are able to simply define the mathematical limit of longevity by the mortality decrease. From the demographic analysis of the historical trends in Switzerland (1876-2001) and Sweden (1861-2001), and the most recent trends in the other eleven developed countries, we confirm that the bending of the shape parameter after characteristic life is correlated with the mortality deceleration (or decrease) and consequently, it is associated with the mathematical limit of longevity. The results suggest that the mathematical limit of longevity can be induced by the mortality deceleration (or decrease) in nature. These findings will give us a breakthrough to study the mortality dynamics at the highest ages.

# Section II. Complementarity principle

**Summary**

We could see that the longevity and the mathematical limit for humans have "complementary" aspects in the Section I. We wish to put forward the complementarity principle on human longevity in this Section. This principle is based upon our researches by the Weon model (the Weibull model with an age-dependent shape parameter) for human survival and mortality curves. In principle for longevity, the shape parameter for humans tends to increase for ages before characteristic life but it tends to decrease for ages after characteristic life, which is attributable to the nature of biological systems to strive to survive healthier and longer. Empirically for ages after characteristic life, the shape parameter shows a quadratic expression, in which the quadratic coefficient indicates the decrease of the shape parameter and it tends to increase with increasing longevity. On the other hand, the quadratic expression is related with the mortality deceleration, plateau, and decrease, inducing the mathematical limit (the mortality rate to be mathematically zero, implying a maximum longevity). Interestingly, the mathematical limit tends to decrease with increasing the quadratic coefficient in the Section I. We are able to conclude that the mathematical limit decreases as the longevity increases. From the phenomenological explanations about the complementarity on longevity and the analysis of the demographic data for Switzerland (1876-2002), we come to the conclusion that there would be the complementarity between the longevity and the mathematical limit. The complementarity principle suggests that the human longevity can be limited.

## 1. Introduction

In our researches, we introduced, developed and established a new concept, model and methodology for studying human longevity in terms of demographic basis. We call the new model the "Weon model", which is a general model modified from the Weibull model with an age-dependent shape parameter to describe human survival and mortality curves. The Weon model is completely different from the Weibull model in terms of the 'age dependence of the shape parameter'. The shape parameter for humans is a function of age is valid for the typical human survival and mortality curves with a certain degree of universality in many countries. We evaluate the age dependence of the shape parameter to determine an adequate mathematical expression of the shape parameter, after determination of the value of the characteristic life graphically in the survival curve. The age-dependent shape parameter distinguishes humans (Weon model) from technical devices (Weibull model).

By the way, the latest research left an interesting question for us. According to the Section I, the trends of the mathematical limit of longevity in recent decades seem to be contrary to all our knowledge. We note that the characteristic life in both of Switzerland and Sweden has increased constantly for more than a century, which is consistent with other literature (for instance, Wilmoth et al. 2000; Oeppen and Vaupel 2002; Robine and Vaupel 2002). On the contrary, the mathematical limits of longevity seem to tend to decrease in recent decades. Furthermore, the mathematical limits of longevity in the thirteen developed countries approach approximately 120 to 130 years in recent. Interestingly, the bending of the shape parameter tends to be more intensive with increasing characteristic life in the Section I. This result suggests that there probably is an inherent correlation between the mortality dynamics and the increase of characteristic life. The mathematical limit of longevity may be associated with the fundamental principle of human longevity. We need a consistent answer for this.

In this Section, we wish to reconsider the origin of the age dependence of the shape parameter in the Weon model and the recent paradoxical trends of the longevity and the mathematical limit. As a consequence, we find out the complementarity principle on human longevity. From the phenomenological explanations about the complementarity on longevity and the analysis of the demographic data for Switzerland (1876-2002), we come to the conclusion that there would be the complementarity between the longevity and the mathematical limit.

## 2. Complementarity between longevity and limit: phenomenological explanations

i) <u>A reason for longevity</u>: The essence of the Weon model is the age dependence of the shape parameter. What is the origin of the age-dependent shape parameter? In the Weon model, the characteristic life ($\alpha$) in the survival curves has a unique feature, which is that all survival curves pass through the characteristic life. So, if the shape parameter is constant with age, a higher value of the survival probability ($S$) at ages before characteristic life ($t < \alpha$) becomes a lower value of the survival probability ($S$) at ages after characteristic life ($t > \alpha$). In principle for the highest value of $S$ or for longevity at all times, the shape parameter ($\beta(t)$) should be variable according to the characteristic life ($\alpha$) as shown in Fig. 10; "for longevity, $\beta(t)$ increases at $t < \alpha$ but it decreases at $t > \alpha$." Fig. 11 shows a descriptive picture of evolution of $\beta(t)$ for longevity. This is attributable to the nature of biological systems to strive to survive healthier and longer. This is a consistent reason for longevity. According to the Weon model, empirically the shape parameter shows a quadratic expression, $\beta(t) = \beta_0 + \beta_1 t + \beta_2 t^2$, at $t > \alpha$, in which the quadratic coefficient ($\beta_2$) indicates the decrease of $\beta(t)$ at $t > \alpha$. Thus, the longevity tends to increase with increasing $\beta_2$.

ii) <u>A reason for limit of longevity</u>: In the Section I, we could see that the mortality dynamics (deceleration, plateau, and decrease) are a consequence of the decrease of the shape parameter at $t > \alpha$. According to the Weon model, the quadratic expression is obviously related with the mortality deceleration, plateau, and decrease, inducing the mathematical limit (the mortality rate to be mathematically zero, implying a maximum longevity). Interestingly, the mathematical limit tends to decrease with increasing quadratic coefficient ($\beta_2$) in the Section I. Consequently, we are able to conclude that the mathematical limit decreases as the longevity increases, which shows "complementary" aspects.

The above phenomenological explanations suggest that there would be the complementarity between the longevity and the mathematical limit as shown in Fig. 12. It is very interesting that the reason for longevity, especially in terms of the decrease of the shape parameter for ages after characteristic life, may be the reason for limit of longevity in nature. In the Section I, we could see that the mathematical limit approaches around ages 120-130, which can be considered to be a fundamental limit of longevity.

## 3. Demographic evidence for Switzerland

In the Section I, we analyzed in detail the demographic data for Switzerland (1876-2001). In recent we obtained the latest demographic data for Switzerland (2002) from the Human Mortality Database (available at http://www.mortality.org). So, we combine the data to demonstrate the demographic evidence for Switzerland over time from 1876 up to 2002.

The age dependence of the shape parameter intrinsically makes the mortality function complex and difficult for modeling. Roughly a linear expression is appropriate for ages 0-80 (Weon 2004a, 2004b). For the best fits to the demographic data over the total life span; a cubic or a quartic expression is appropriate for ages 0-20, a linear or a quadratic expression is appropriate for ages 20-80 and a quadratic expression is appropriate for ages 80+. We demonstrate an example of modeling the mortality rate through the Weon model for Switzerland (2002) in Fig. 13. The best fits for $\beta(t)$ in Switzerland (2002) are given as follows; $1.09605 + 0.1680t - 0.00385t^2 + 0.00006t^3$ ($r^2 = 0.9992$) for ages 0-20, $1.05695 + 0.09705t$ ($r^2 = 0.9975$) for ages 20-80 and $-25.08726 + 0.72140 - 0.00365t^2$ ($r^2 = 0.9559$) for ages 80-109.

Specifically, the mortality curves for higher ages (110+) are important to understand the longevity. According to the Weon model, the quadratic expression for ages 80-109 is valid with a certain degree of university in many modern developed countries, which enables us to predict that the mortality rate decreases after a plateau around ages 110-115 and the mathematical limit of longevity emerges around ages 120-130, for example for Switzerland (2002), as shown in Fig. 14. If the quadratic expression is valid for ages 110+, we are able to predict the mortality rate at the highest ages. This pattern of the mortality dynamics at the highest ages by the Weon model is consistent with the other assertions (for instance, Vaupel et al. 1998; Robine and Vaupel 2002; Helfand and Inouye 2002).

In the Section I, we successfully used the quadratic expression for the description of the shape parameter after ages 80+. Interestingly, the quadratic coefficient is important to evaluate the mathematical limit of longevity, since it determines the slope with age in the derivative of the quadratic expression of the shape parameter. We are able to see that for longevity, $\beta(t)$ increases at $t < \alpha$ but it decreases at $t > \alpha$ in Fig. 15. We are also able to see that the mathematical limit decreases as the characteristic life increases in Fig. 16.

## 4. Ultimate value of mathematical limit for Switzerland and Sweden

In Fig. 16, the trend line for the mathematical limit is estimated to be $t = 123.9 + 161.2 \times \exp(2308 \times \beta_2)$ ($r^2 = 0.9750$) and the trend line for the longevity tendency (characteristic life) is estimated to be $t = 62.58 - 6033 \times \beta_2$ ($r^2 = 0.7307$). Note that $\beta_2$ is negative in this case. Especially, we are able to predict the ultimate value of the mathematical limit as '$\beta_2 \to -\infty$' to be approximately 123.9 years for Switzerland, which may be the ultimate limit of longevity for Switzerland. In this way, we are able to analyze the case of Sweden (1861-2002) as well as that of Switzerland (1876-2002) to estimate the trend line of the mathematical limit for Switzerland and Sweden to be $t = 122.7 + 161.1 \times \exp(2271 \times \beta_2)$ ($r^2 = 0.9268$) in Fig. 17. In this case, the ultimate value of the mathematical limit as '$\beta_2 \to -\infty$' is predicted to be approximately 122.7 years for Switzerland and Sweden. The cases of Switzerland and Sweden are likely to be largely similar to that of the other developed countries. Interestingly, this ultimate value for longevity is approximate to the world record for human longevity in the case of Jeanne Calment, who lived to the ripe old age of 122 years and 164 days (122.45 years).

## 5. Conclusions

In this Section, we put forward the complementarity principle on human longevity that the longevity and the mathematical limit for humans have "complementary" aspects. This principle is the natural and logical consequence of our researches by the Weon model (the Weibull model with an age-dependent shape parameter). In principle for longevity, the shape parameter for humans tends to increase for ages before characteristic life but it tends to decrease for ages after characteristic life, which is attributable to the nature of biological systems to strive to survive healthier and longer. Empirically for ages after characteristic life, the shape parameter shows a quadratic expression, in which the quadratic coefficient indicates the decrease of the shape parameter and it tends to increase with increasing longevity. On the other hand, the quadratic expression is related with the mortality deceleration, plateau, and decrease, which could induce a mathematical limit (the mortality rate to be mathematically zero, implying a maximum longevity). Interestingly, the mathematical limit tends to decrease with increasing the quadratic coefficient. From the phenomenological explanations about the complementarity on longevity and the analysis of the demographic data for Switzerland (1876-2002), we come to the conclusion that there would be the complementarity between the longevity and the mathematical limit. Consequently, the human longevity is likely to be limited by the complementarity principle.

**Figures and Tables for the Section I.**

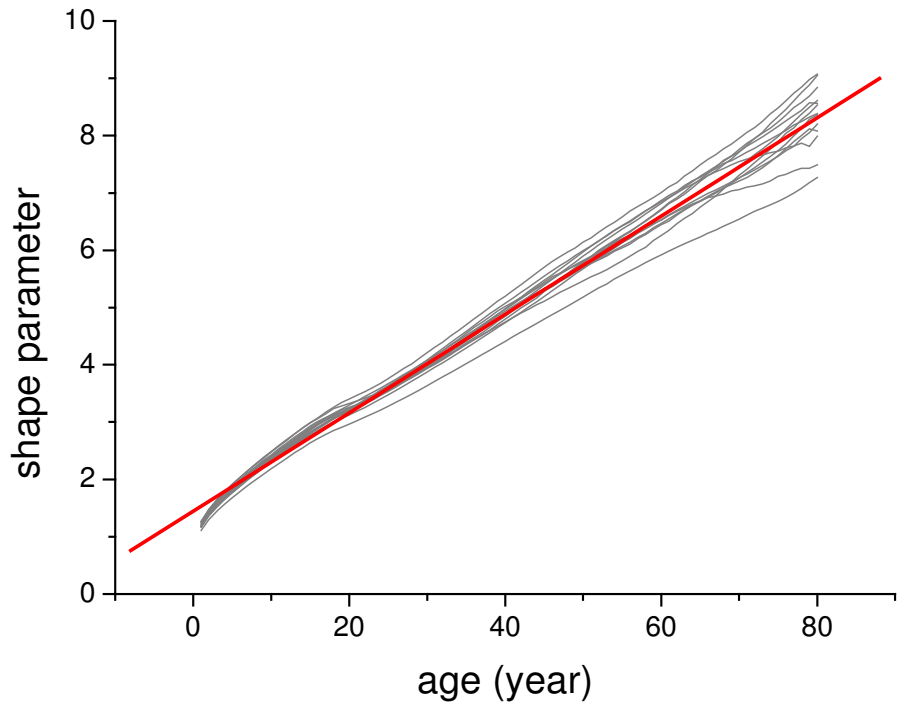

Fig. 1A. Universality of linear progression of the shape parameter for ages 0-80 in thirteen developed countries (1996-2001).

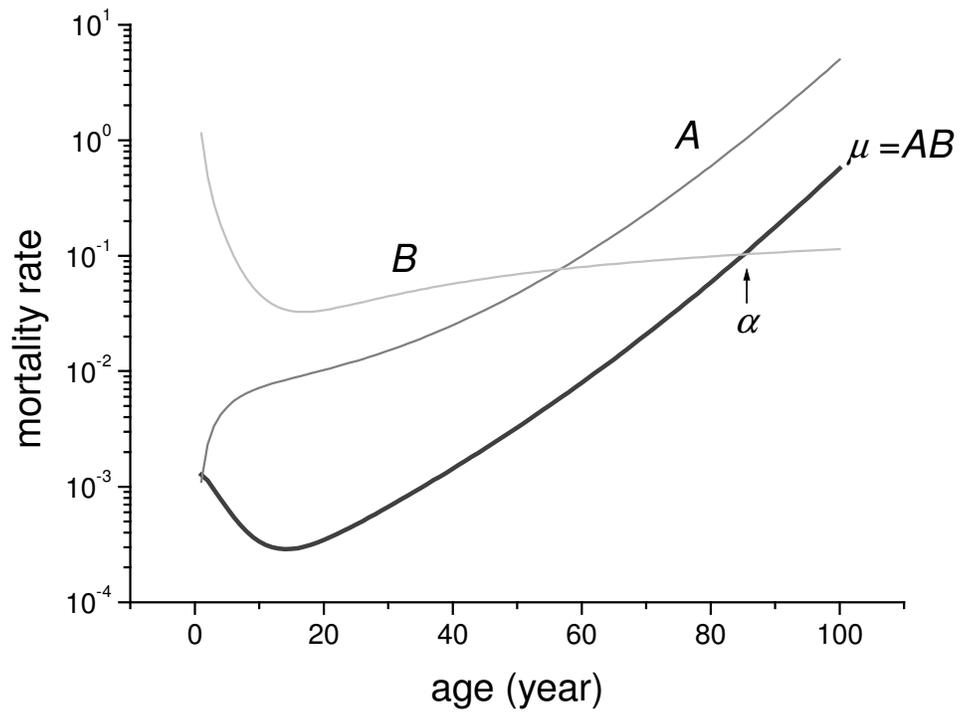

Fig. 1B. Simulation of the mortality rate through the universality of the shape parameter.

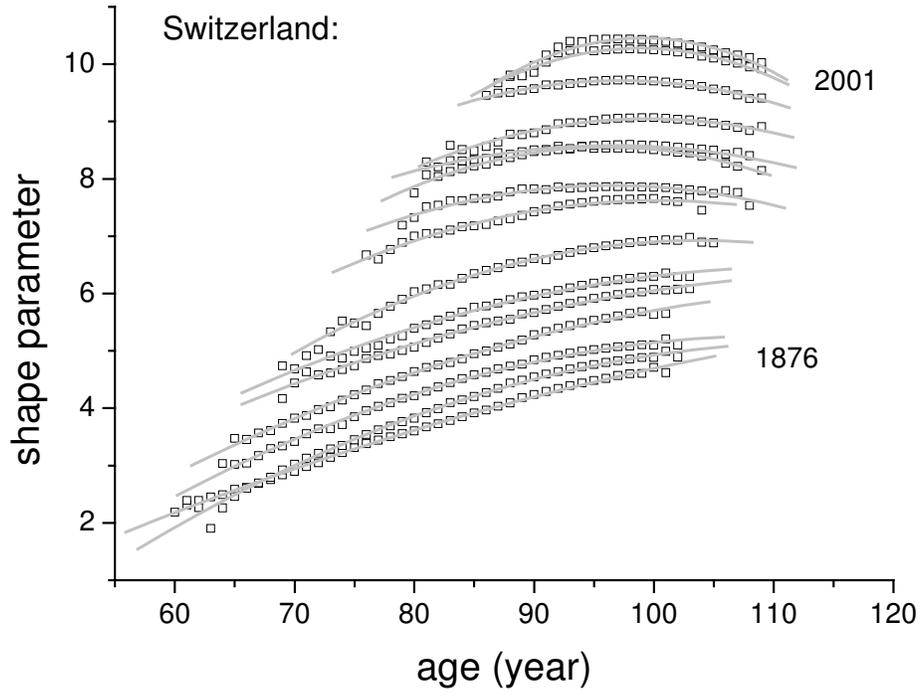

Fig. 2A. Historical trends of the shape parameters after characteristic life in Switzerland (1876-2001).

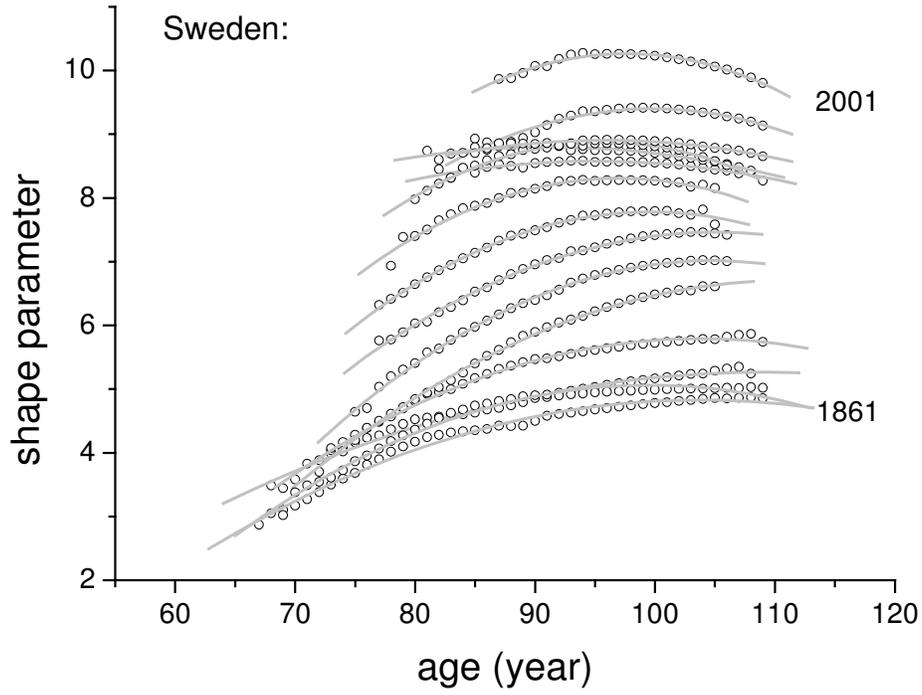

Fig. 2B. Historical trends of the shape parameters after characteristic life in Sweden (1861-2001).

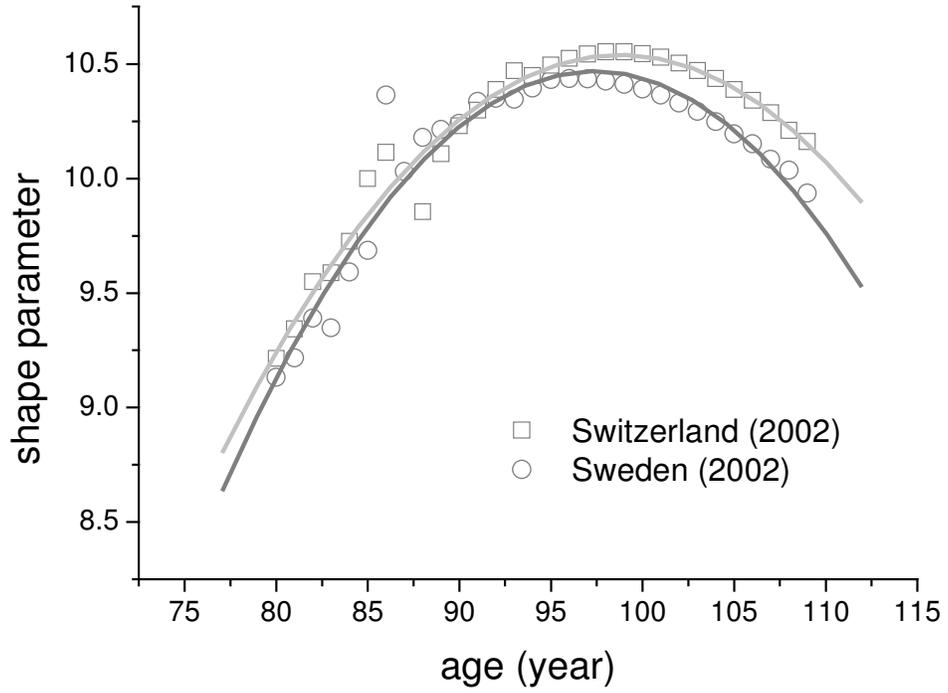

Fig. 3A. Modeling the shape parameter as a quadratic expression for Switzerland (2002) and Sweden (2002).

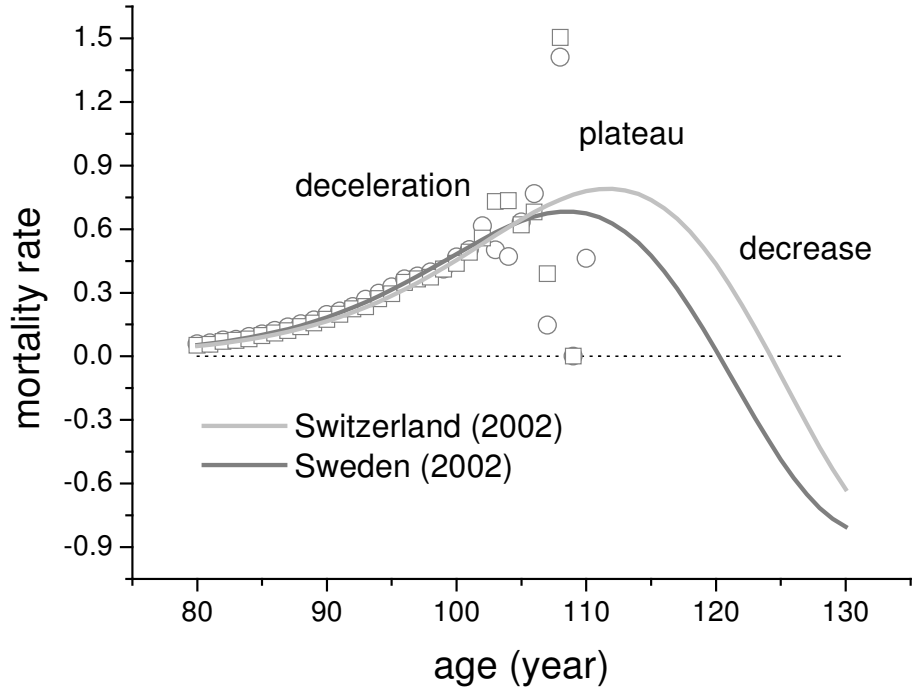

Fig. 3B. Modeling the mortality rate through modeling the shape parameter and the death rate (mortality) data for Switzerland (2002) and Sweden (2002).

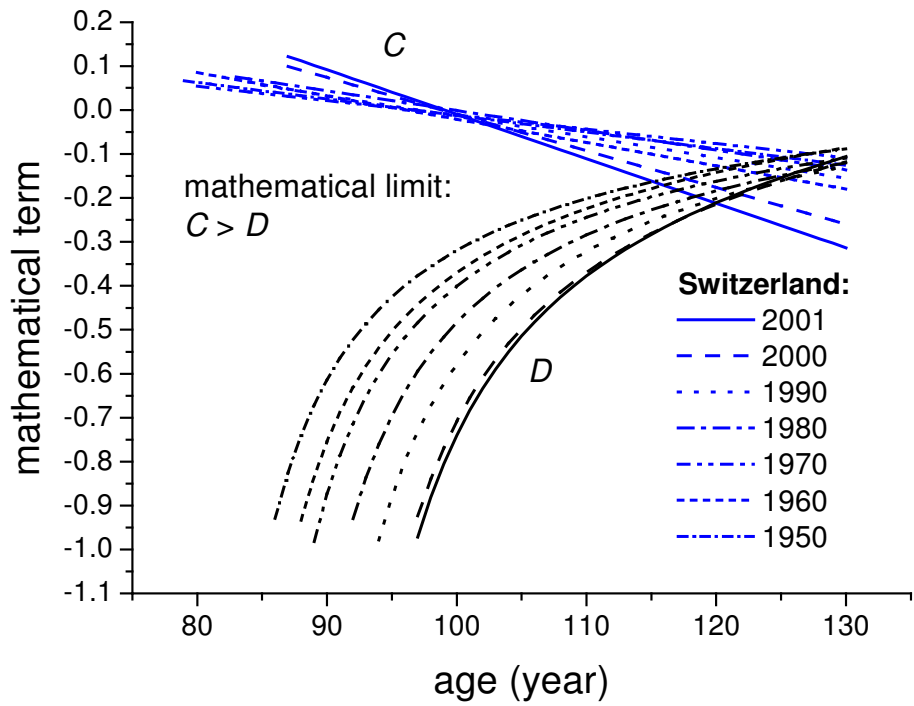

Fig. 4A. Calculations of mathematical terms, *C* and *D*, as a function of age for Switzerland.

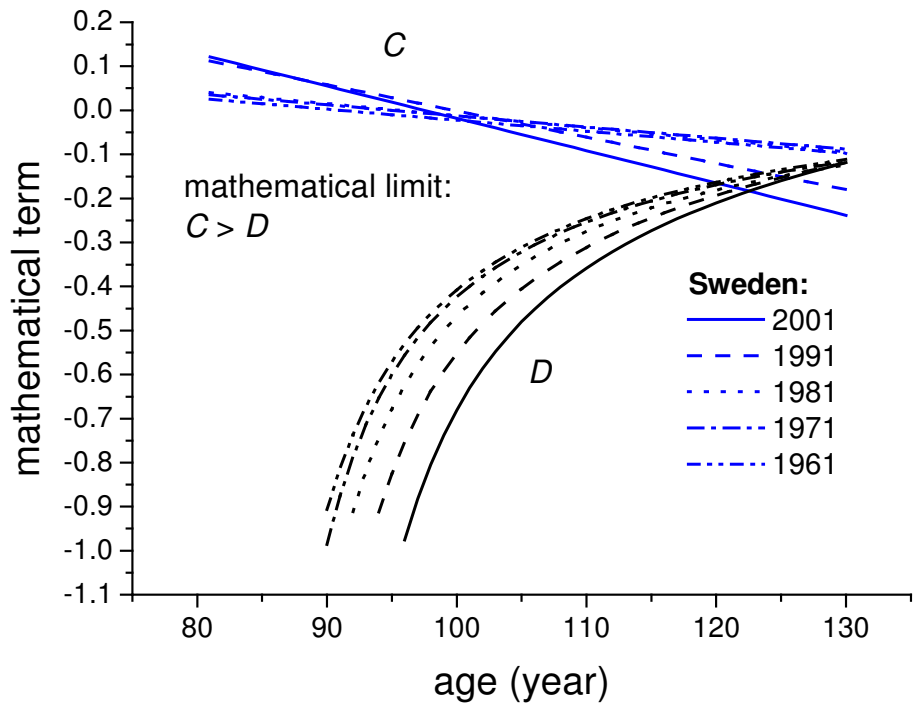

Fig. 4B. Calculations of mathematical terms, $C$ and $D$, as a function of age for Sweden.

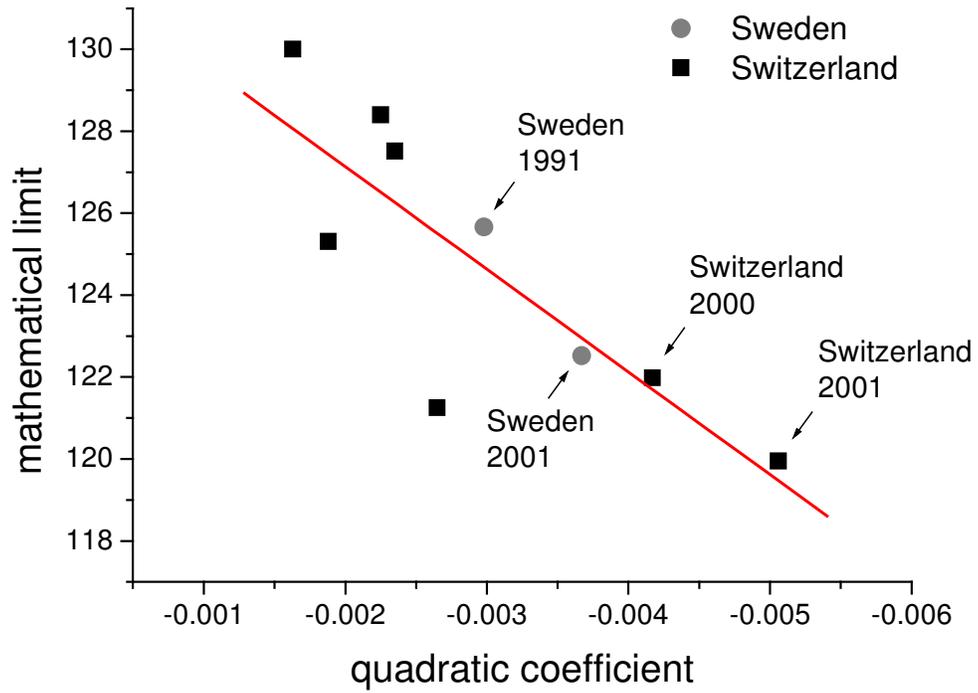

Fig. 5. Shortening mathematical limits of longevity with increasing the quadratic coefficient ($\beta_2$) in both of Switzerland and Sweden.

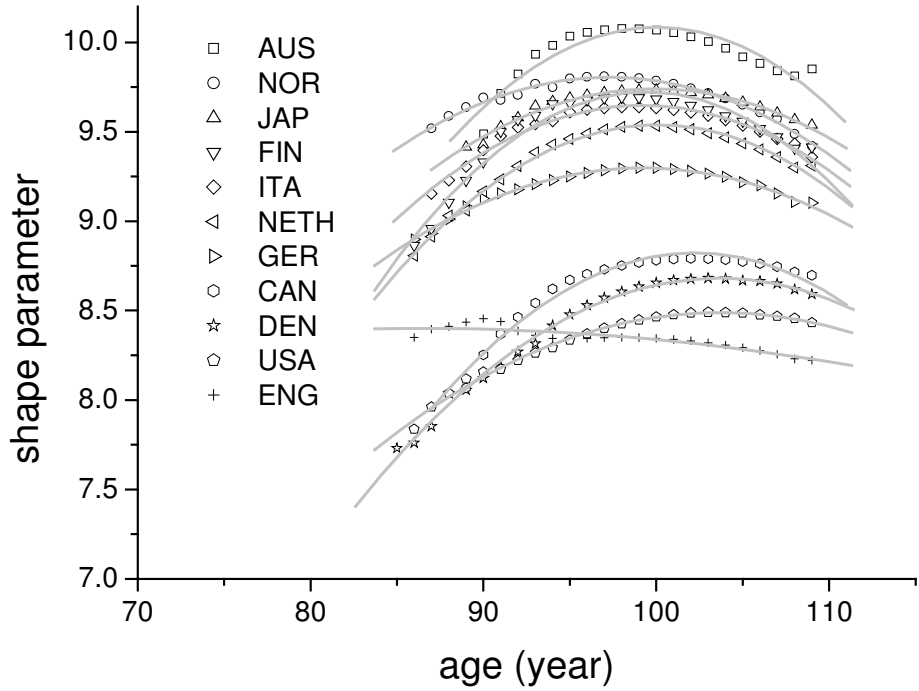

Fig. 6. Trends of the shape parameters after characteristic life in the other eleven developed countries (1996-2001) as the most recent trends. Note: AUS-Austria, NOR-Norway, JAP-Japan, FIN-Finland, ITA-Italy, NETH-Netherlands, GER-Germany, CAN-Canada, DEN-Denmark, USA-Unites State of America and ENG-England & Wales.

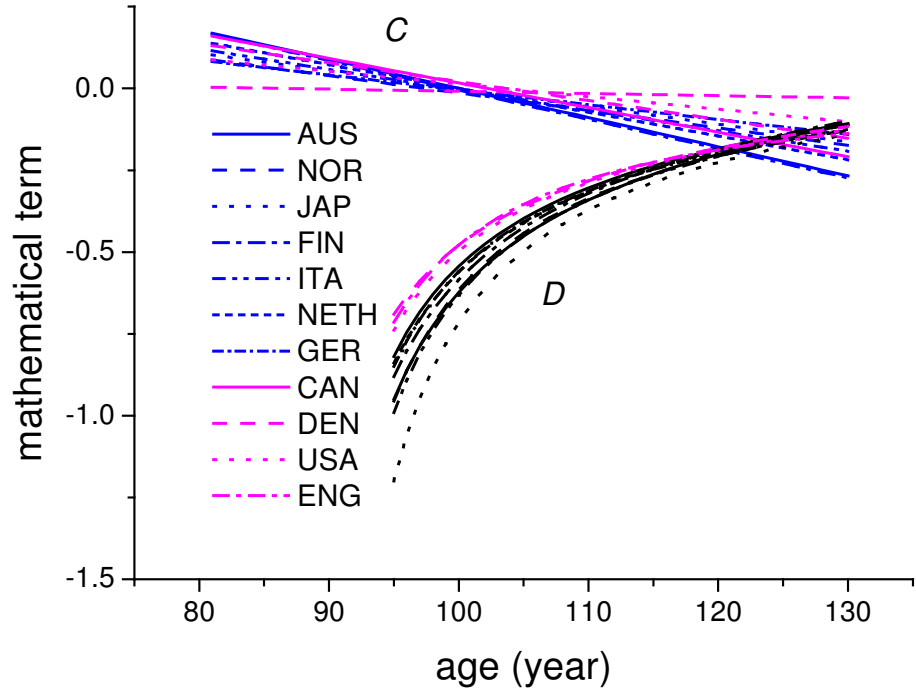

Fig. 7. Calculations of mathematical terms, *C* and *D*, as a function of age in the eleven developed countries. Note: AUS-Austria, NOR-Norway, JAP-Japan, FIN-Finland, ITA-Italy, NETH-Netherlands, GER-Germany, CAN-Canada, DEN-Denmark, USA-Unites State of America and ENG-England & Wales.

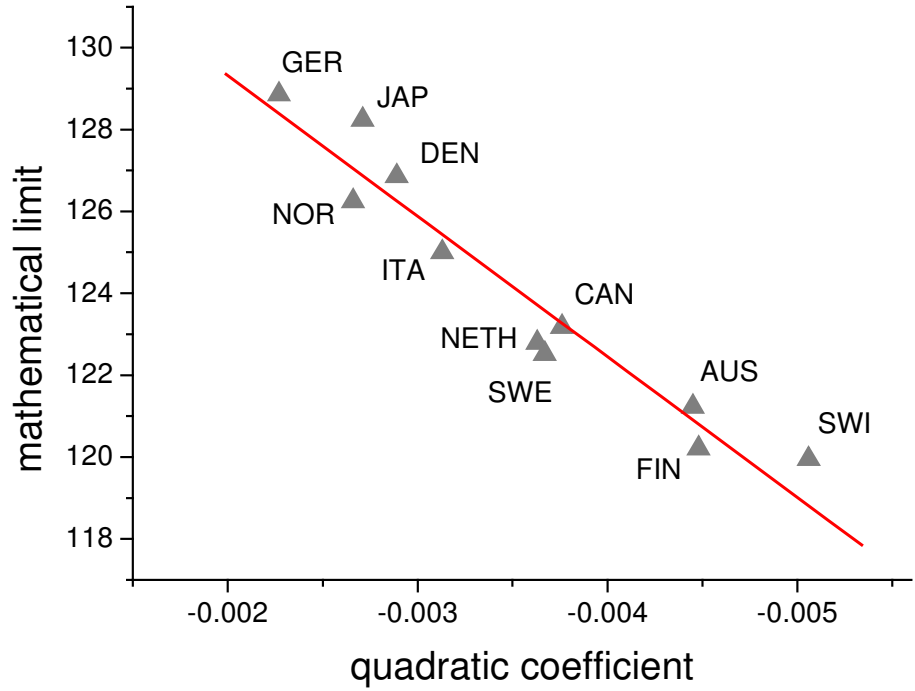

Fig. 8. Shortening mathematical limits of longevity with increasing the quadratic coefficient ($β_2$) in the eleven developed countries including Switzerland and Sweden. Note: AUS-Austria, NOR-Norway, JAP-Japan, FIN-Finland, ITA-Italy, NETH-Netherlands, GER-Germany, CAN-Canada, DEN-Denmark, USA-Unites State of America and ENG-England & Wales.

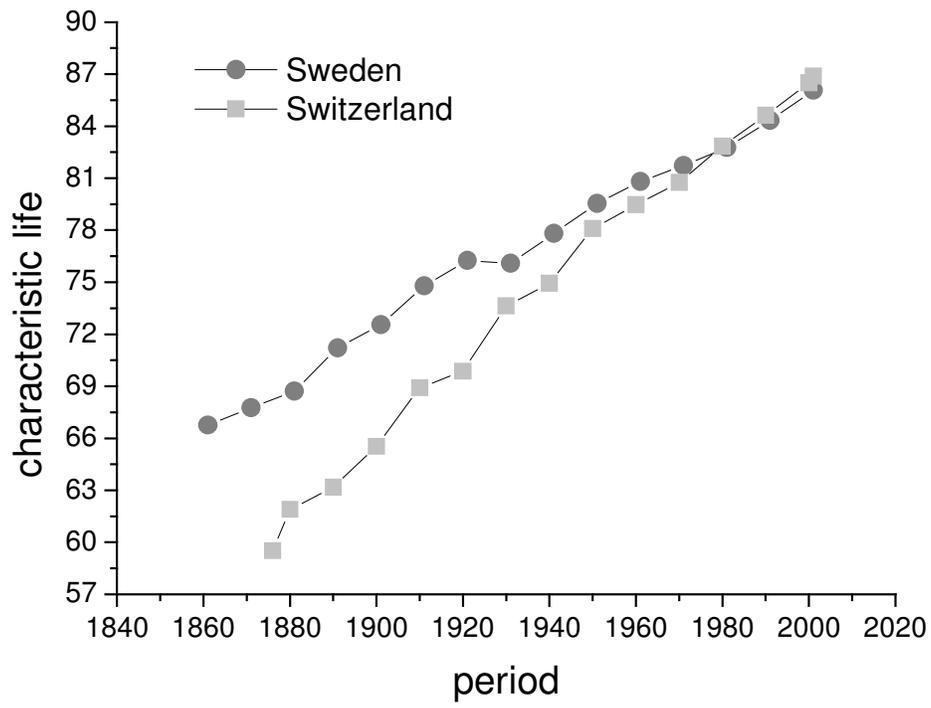

Fig. 9. Historical trends of characteristic life in both of Switzerland and Sweden.

Table 1. Shape parameter estimated as a quadratic expression after characteristic life for 1876-2001 for Switzerland.

| Year | $\beta_0$ | $\beta_1$ | $\beta_2$ | $r^2$ | (P-value) |
|------|-----------|-----------|-----------|---------|-----------|
| 2001 | -39.22154 | 1.00226 | -0.00506 | 0.90494 | (<0.0001) |
| 2000 | -30.49174 | 0.82469 | -0.00417 | 0.93217 | |
| 1990 | -12.42385 | 0.45586 | -0.00235 | 0.97375 | |
| 1980 | -13.29566 | 0.44890 | -0.00225 | 0.93770 | |
| 1970 | -6.617620 | 0.31492 | -0.00163 | 0.84224 | |
| 1960 | -15.90502 | 0.50906 | -0.00265 | 0.95876 | |
| 1950 | -9.649830 | 0.36351 | -0.00188 | 0.92426 | |
| 1940 | -9.475550 | 0.34073 | -0.00170 | 0.97375 | |
| 1930 | -11.57780 | 0.35689 | -0.00172 | 0.99198 | |
| 1920 | -6.145470 | 0.22396 | -0.00099 | 0.97999 | |
| 1910 | -4.798780 | 0.18597 | -0.00077 | 0.98526 | |
| 1900 | -6.486440 | 0.20651 | -0.00085 | 0.99744 | |
| 1890 | -8.211080 | 0.24414 | -0.00111 | 0.99660 | |
| 1880 | -9.158820 | 0.25042 | -0.00110 | 0.99056 | |
| 1876 | -4.318390 | 0.13551 | -0.00045 | 0.99815 | |

Note: $\beta(t) = \beta_0 + \beta_1 t + \beta_2 t^2$

Table 2. Shape parameter estimated as a quadratic expression after characteristic life for 1861-2001 for Sweden.

| Year | $\beta_0$ | $\beta_1$ | $\beta_2$ | $r^2$ | (P-value) |
|------|-----------|-----------|-----------|---------|-----------|
| 2001 | -24.68505 | 0.71594 | -0.00367 | 0.96717 | (<0.0001) |
| 1991 | -20.28298 | 0.59479 | -0.00298 | 0.91251 | |
| 1981 | -3.653530 | 0.26206 | -0.00137 | 0.88759 | |
| 1971 | -2.792430 | 0.23941 | -0.00126 | 0.84286 | |
| 1961 | -1.574590 | 0.22763 | -0.00125 | 0.91608 | |
| 1951 | -22.55137 | 0.65682 | -0.00343 | 0.98771 | |
| 1941 | -21.82020 | 0.62146 | -0.00320 | 0.97352 | |
| 1931 | -21.50172 | 0.58798 | -0.00295 | 0.99537 | |
| 1921 | -17.87890 | 0.48209 | -0.00229 | 0.99670 | |
| 1911 | -22.02449 | 0.55444 | -0.00265 | 0.99770 | |
| 1901 | -17.00765 | 0.42768 | -0.00193 | 0.99772 | |
| 1891 | -14.30066 | 0.38774 | -0.00187 | 0.99151 | |
| 1881 | -14.62550 | 0.39595 | -0.00199 | 0.98411 | |
| 1871 | -6.756830 | 0.22012 | -0.00101 | 0.98537 | |
| 1861 | -10.20184 | 0.28998 | -0.00140 | 0.98905 | |

Note: $\beta(t) = \beta_0 + \beta_1 t + \beta_2 t^2$

Table 3. Shape parameter estimated as a quadratic expression after characteristic life for the eleven developed countries.

| Country | $\beta_0$ | $\beta_1$ | $\beta_2$ | $r^2$ | (P-value) |
|---|---|---|---|---|---|
| Austria | -34.43938 | 0.89024 | -0.00445 | 0.84619 | (<0.0001) |
| Norway | -15.38753 | 0.51811 | -0.00266 | 0.98490 | |
| Japan | -17.30636 | 0.54138 | -0.00271 | 0.97326 | |
| Finland | -34.59301 | 0.89137 | -0.00448 | 0.95626 | |
| Italy | -21.20022 | 0.62189 | -0.00313 | 0.98166 | |
| Netherlands | -26.79699 | 0.72611 | -0.00363 | 0.99124 | |
| Germany | -13.02618 | 0.44991 | -0.00227 | 0.99067 | |
| Canada | -30.47682 | 0.76892 | -0.00376 | 0.96768 | |
| Denmark | -22.35540 | 0.59909 | -0.00289 | 0.99459 | |
| USA | -12.23463 | 0.39989 | -0.00193 | 0.99228 | |
| England & Wales | +5.92728 | 0.05735 | -0.00033 | 0.82539 | |

Note: $\beta(t) = \beta_0 + \beta_1 t + \beta_2 t^2$

**Figures for the Section II.**

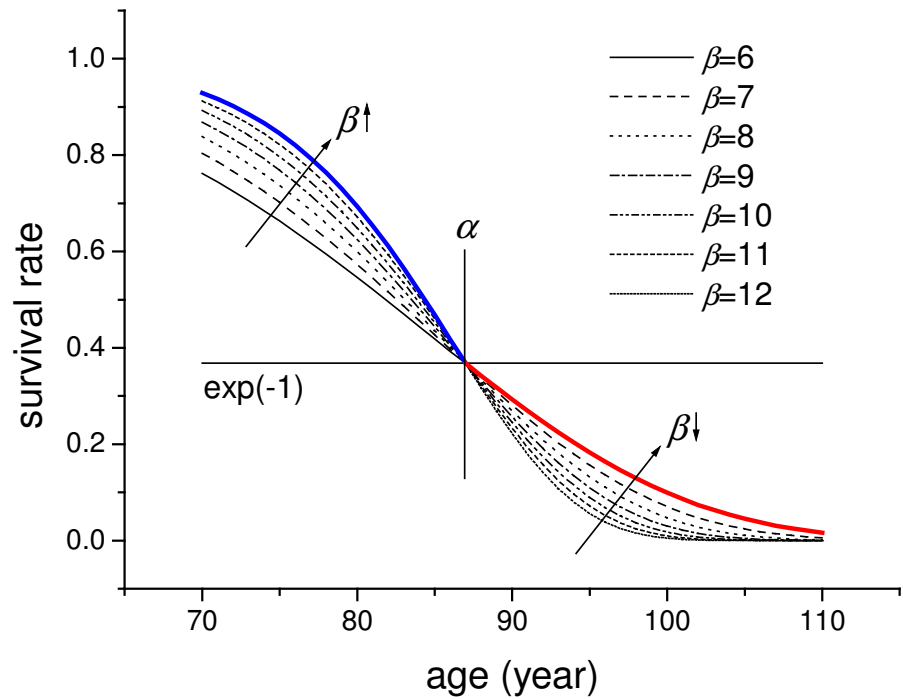

Fig. 10. Behavior of the shape parameter behavior for longevity before and after characteristic life.

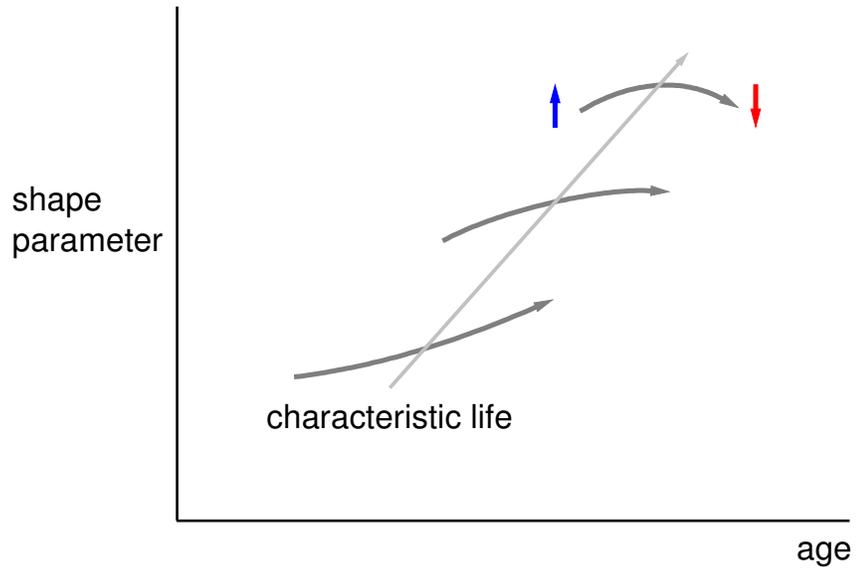

Fig. 11. Behavior of the shape parameter for longevity with increasing characteristic life.

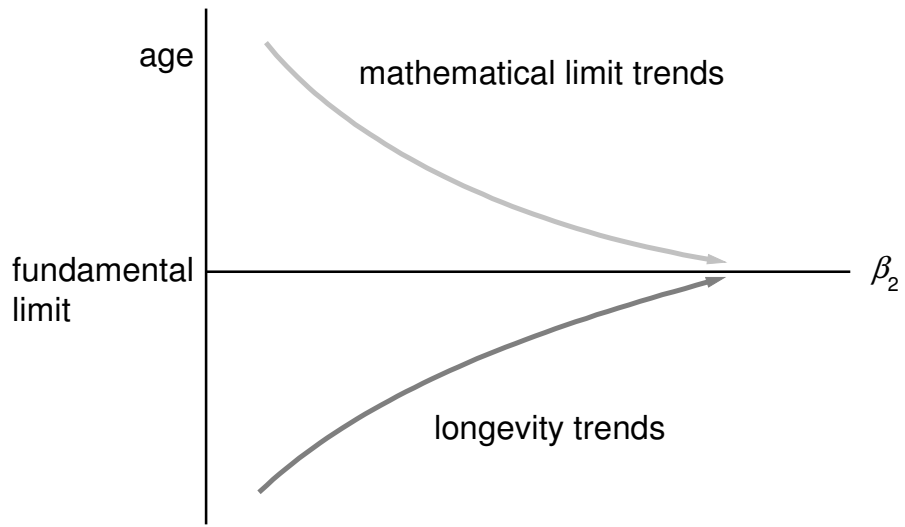

Fig. 12. Complementarity between trends of the longevity and the mathematical limit.

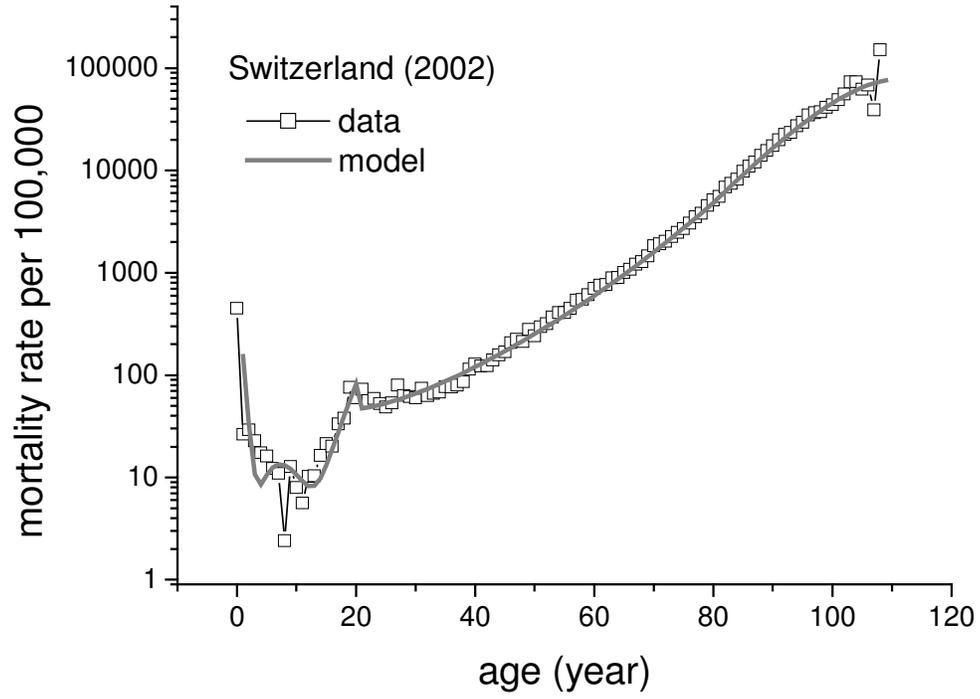

Fig. 13. An example of modeling the mortality rate for Switzerland (2002).

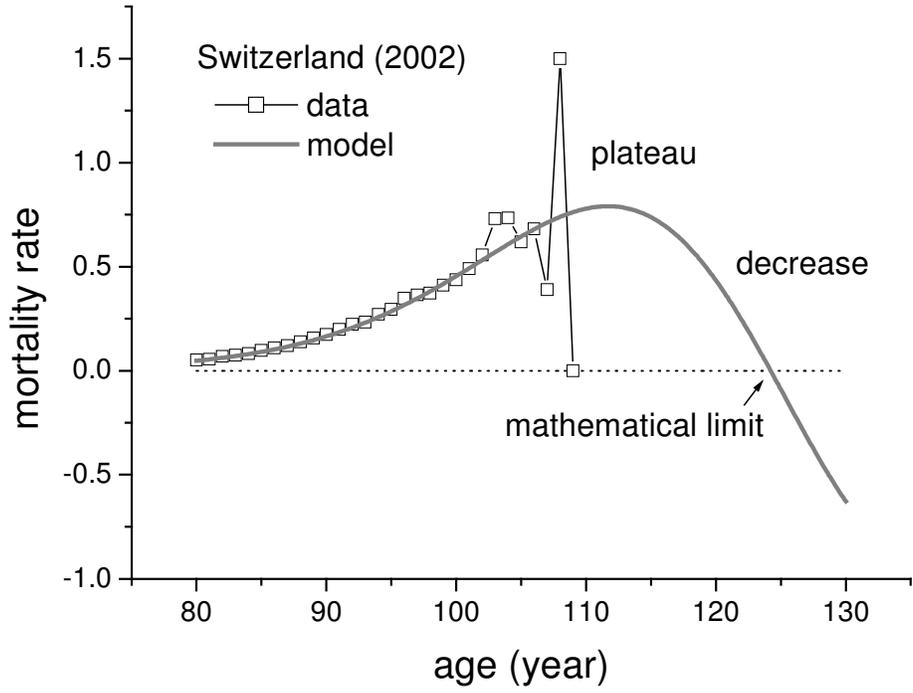

Fig. 14. An example of modeling the mortality rate at the highest ages for Switzerland (2002).

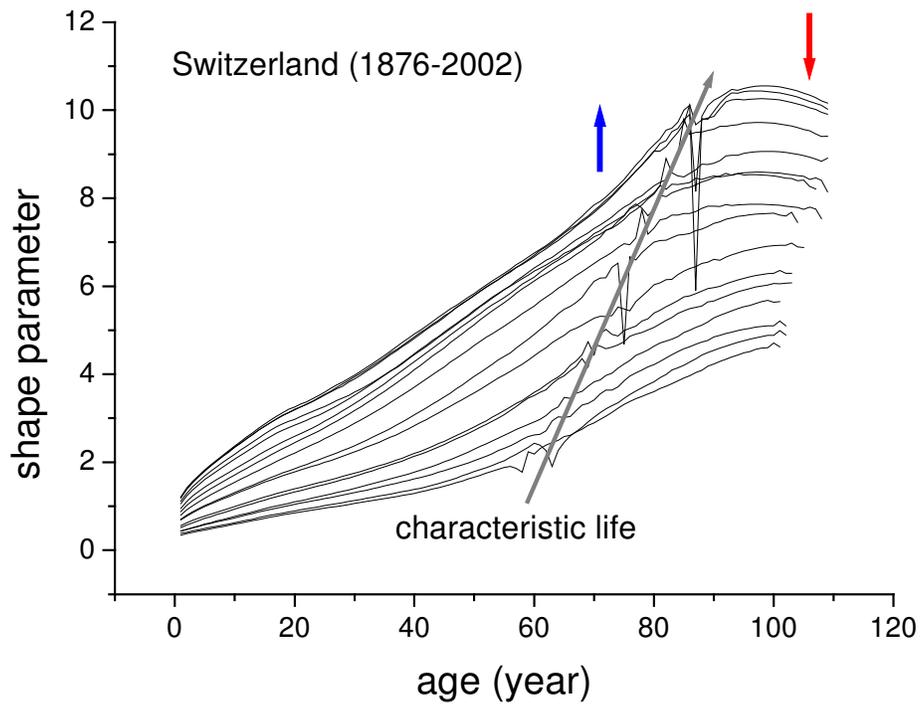

Fig. 15. Demographic evidence of the shape parameter behavior with increasing characteristic life for Switzerland (1876-2002).

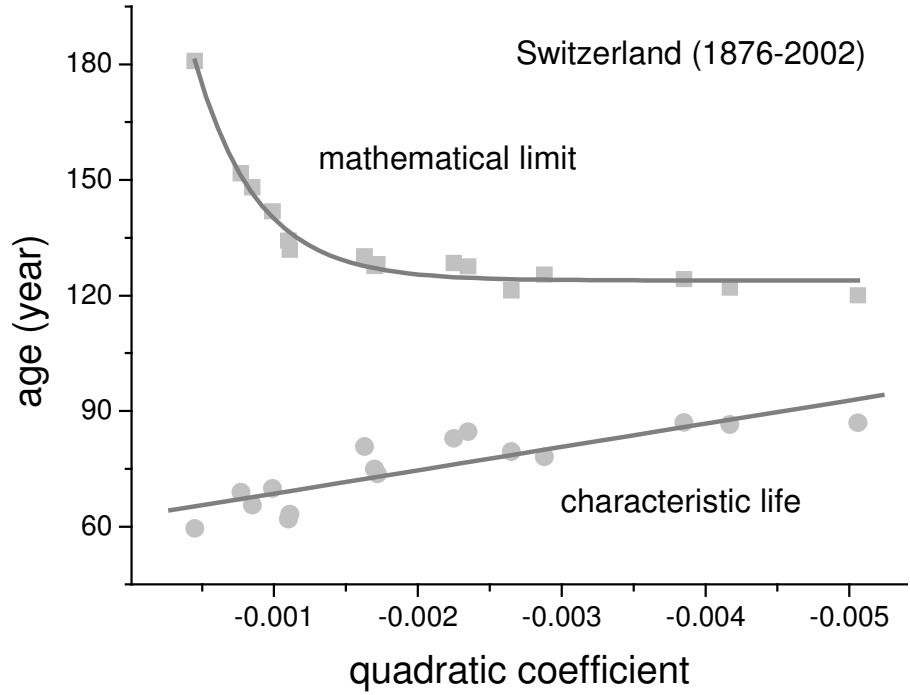

Fig. 16. Demographic evidence of the complementarity between the longevity (characteristic life) and the mathematical limit with increasing the quadratic coefficient for Switzerland (1876-2002).

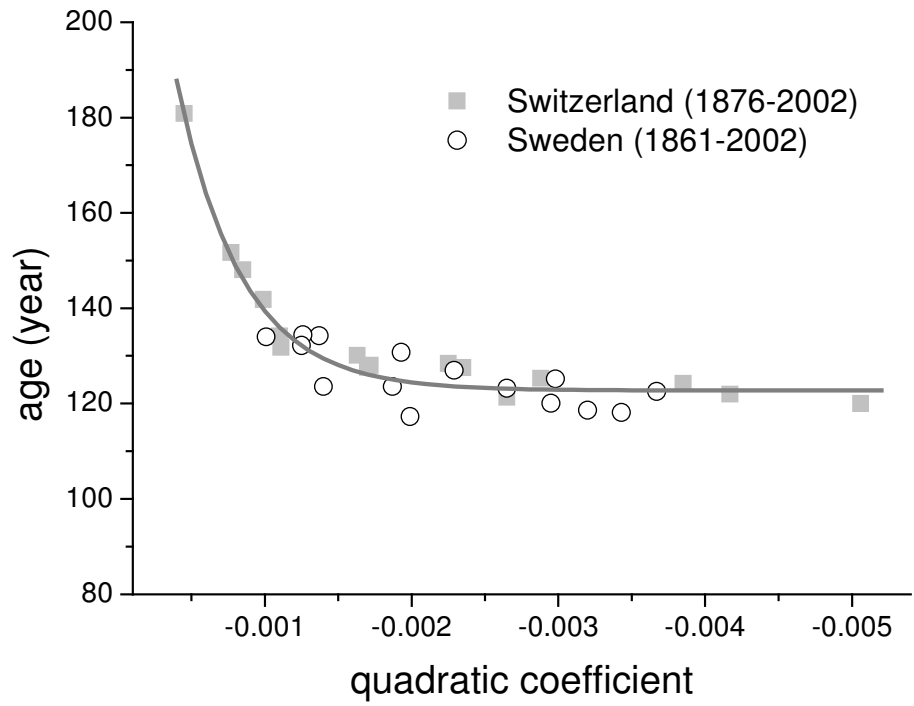

Fig. 17. Demographic evidence of the trends of the mathematical limit with increasing the quadratic coefficient for Switzerland (1876-2002) and Sweden (1861-2002).